\documentclass[twocolumn,aps,prl,superscriptaddress]{revtex4}
\usepackage{amssymb}
\usepackage{amsmath}
\usepackage{graphicx} 
\usepackage{color} 
\newcommand{\ket}[1]{\left\vert#1\right\rangle}
\begin{document}
\title{Experimental Realization of a Controlled-NOT Gate with Four-Photon Six-Qubit Cluster States}
\author{Wei-Bo Gao}
\affiliation{Hefei National Laboratory for Physical Sciences at
Microscale and Department of Modern Physics, University of Science
and Technology of China, Hefei, Anhui 230026, China}
\author{Ping Xu}
\affiliation{Hefei National Laboratory for Physical Sciences at
Microscale and Department of Modern Physics, University of Science
and Technology of China, Hefei, Anhui 230026, China}
\author{Xing-Can Yao}
\affiliation{Hefei National Laboratory for Physical Sciences at
Microscale and Department of Modern Physics, University of Science
and Technology of China, Hefei, Anhui 230026, China}
\author{Otfried G\"{u}hne}
\affiliation{Institut f\"{u}r Quantenoptik und Quanteninformation,
\"{O}sterreichische Akademie der Wissenschaften, Technikerstra{\ss}e
21A, A-6020 Innsbruck, Austria\\} \affiliation{Institut f{\"u}r
theoretische Physik, Universit{\"a}t Innsbruck, Technikerstra{\ss}e
25, A-6020 Innsbruck, Austria\\}
\author{Ad\'{a}n Cabello}
\affiliation{Departamento de F\'{\i}sica Aplicada II, Universidad de
Sevilla, E-41012 Sevilla, Spain\\}
\author{Chao-Yang Lu}
\affiliation{Hefei National Laboratory for Physical Sciences at
Microscale and Department of Modern Physics, University of Science
and Technology of China, Hefei, Anhui 230026, China}

\author{Cheng-Zhi Peng}
\affiliation{Hefei National Laboratory for Physical Sciences at
Microscale and Department of Modern Physics, University of Science
and Technology of China, Hefei, Anhui 230026, China}
\author{Zeng-Bing Chen}
\affiliation{Hefei National Laboratory for Physical Sciences at
Microscale and Department of Modern Physics, University of Science
and Technology of China, Hefei, Anhui 230026, China}
\author{Jian-Wei Pan}
\affiliation{Hefei National Laboratory for Physical Sciences at
Microscale and Department of Modern Physics, University of Science
and Technology of China, Hefei, Anhui 230026, China}
\affiliation{Physikalisches Institut, Ruprecht-Karls-Universit\"{a}t
Heidelberg, Philosophenweg 12, 69120 Heidelberg, Germany}

\date{\today}


\begin{abstract}
We experimentally demonstrate an optical controlled-NOT (CNOT) gate
with arbitrary single inputs based on a 4-photon 6-qubit cluster
state entangled both in polarization and spatial modes. We first
generate the 6-qubit state, and then, by performing single-qubit
measurements the CNOT gate is applied to arbitrary single input
qubits. To characterize the performance of the gate, we estimate its
quantum process fidelity and prove its entangling capability. In
addition, our results show that the gate cannot be reproduced by
local operations and classical communication. Our experiment shows
that such hyper-entangled cluster states are promising candidates
for efficient optical quantum computation.
\end{abstract}


\pacs{03.67.Lx, 42.50.Dv}

\maketitle


{\em Introduction.---}Cluster states not only provide a useful model
to study multiparticle entanglement \cite{Briegel01,Hein}, but also
have applications in quantum communication \cite{Cleve}, quantum
non-locality \cite{otfried,scarani,Cabello}, and quantum error
correction \cite{Schlingemann}. Specifically, they play a crucial
role in one-way quantum computation \cite{Briegel02}, which is a
promising approach towards scalable quantum computation.
Considerable efforts have been made toward generating and
characterizing multiparticle cluster states, especially in linear
optics \cite{Nielsen,Terry, Duan}. Recently, some 4-photon cluster
states and one-way quantum computation based on them have been
experimentally demonstrated \cite{Walther,Kiesel,yama}. Also, the
6-photon cluster state has been reported \cite{Lu}. An efficient way
to extend the number of qubits without increasing the number of
particles is entangling in various degrees of freedom
\cite{Mair,Kwiat2,Schuck,Matini1,Kai,bangzi,Gao,Benjamin}. This type
of entanglement is called hyper-entanglement \cite{Kwiat1, Kwiat2},
which can have a high generation rate and fidelity, and thus are
particularly suitable for one-way quantum computation \cite{Matini2,
Kai}.

In this paper, we report on the creation of a 4-photon 6-qubit
cluster state entangled in photons' polarization and spatial modes
and an optical controlled-NOT (CNOT) gate with arbitrary
single-qubit inputs based on the state.
To characterize our gate, we obtain an estimation of the quantum
process fidelity and entangling capability
\cite{Hofmann01,Hofmann02}. Moreover, the experimental results show
that quantum parallelism has been achieved in our gate, and thus the
performance of the gate can't be reproduced by local operations and
classical communication \cite{Hofmann02}.




\begin{figure}
[ptb]
\begin{center}
\includegraphics[width=2.2in]%
{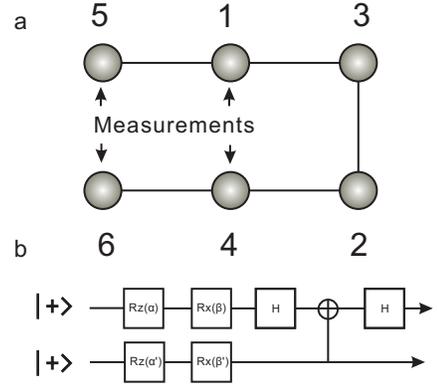}%
\caption{\label{Fig 1}A one-way quantum CNOT gate based on cluster
states. \textbf{a}. 4-photon 6-qubit linear cluster state. Qubits 1,
2, 3, 4 are polarization qubits, and qubits 5, 6 are spatial qubits.
By implementing single-photon measurements and feed-forward
operations depending on the measurement results, the input qubits
are transmitted through a deterministic CNOT gate. \textbf{b}.
Corresponding quantum circuit. $R_z(\alpha)={\rm exp}(i \alpha
Z/2)$, $R_x(\beta)={\rm exp}(i \beta X/2)$, and $H$ denotes a
Hadamard gate.}
\end{center}
\end{figure}


Cluster states are defined as eigenstates of certain sets of local
observables. For intance, an $N$-qubit linear cluster state is the
eigenstate (with eigenvalue $+1$) of the $N$ observables $X_1 Z_2,
Z_1 X_2 Z_3,\ldots, Z_{N-1} X_N$, where $X_i$ and $Z_j$ are Pauli
matrices on the qubits $i$ and $j$, respectively. Given a cluster
state, one-way quantum computation can be performed by making
consecutive single-qubit measurements in the basis $B_k(\alpha)=\{
\ket{\alpha_+}_k, \ket{\alpha_-}_k \}$, where $\ket{\alpha_{\pm}}_k
= (\ket{0}_k \pm e^{i \alpha}\ket{1}_k)/\sqrt{2}$
($\alpha\!\in\!{\mathbb R}$), followed by feed-forward operations
depending on the measurement results.
This measurement basis determines a rotation $R_z(\alpha)={\rm
exp}(i \alpha Z/2)$, followed by a Hadamard operation
$H=(X+Z)/\sqrt{2}$ on the encoded qubits. As depicted in
Fig.~\ref{Fig 1}, based on a linear-type 6-qubit cluster state
$\ket{LC_{6}}$, measurements on qubits 5, 1, 6, 4 in the basis
$\{B_5 (\alpha), B_1 (\beta), B_6 (\alpha'), B_4 (\beta')\}$ will
give an output state on qubits 2, 3 with $(\openone \otimes
H)CNOT[R_x (\beta ')R_z (\alpha ') \otimes HR_x (\beta )R_z (\alpha
)]\left| \tilde{0} \right\rangle_2 \left| \tilde{0} \right\rangle_3
$, where $\left|\tilde{0}\right\rangle=\frac{1}{\sqrt{2}}(\left| 0
\right\rangle+\left| 1 \right\rangle)$. $R_x (\beta ')R_z (\alpha
')$ and $HR_x (\beta )R_z (\alpha )$ are sufficient to realize
arbitrary single-qubit rotations; thus, after compensating the $H$
gate behind the CNOT gate, a CNOT gate with arbitrary single-qubit
inputs can be achieved. Qubits 2 and 3 are, respectively, the
control and target qubits.

\begin{figure}
 \includegraphics[width=7cm]{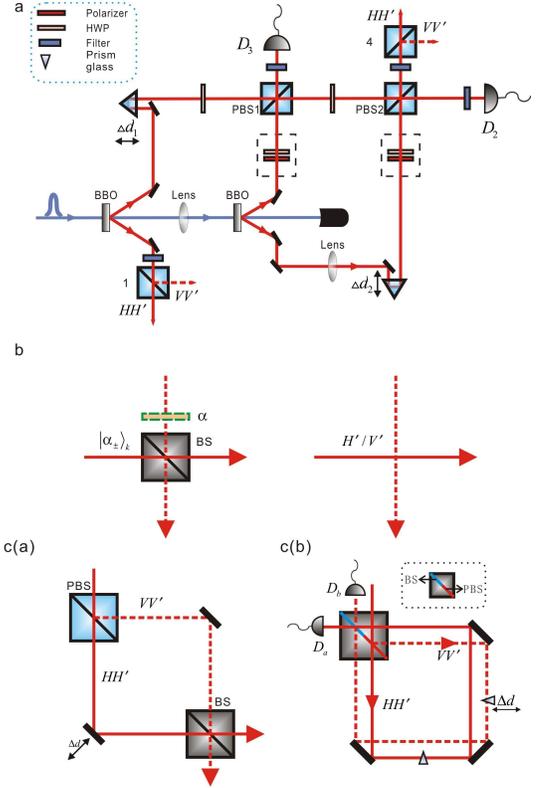}\\
 \caption{\label{Fig 2}Schematic of the experimental setup. \textbf{a}. The setup to generate the required entanglement state.
 Femtosecond laser pulses ($\approx$ 200 fs, 76 MHz, 788nm) are
 converted to ultraviolet pulses through a frequency doubler $LiB_{3}O_{5}$
 (LBO) crystal (not shown). The pulses go through two main $\beta$-barium borate (BBO) crystals (2mm), generating two pairs of photons.
 The observed two-fold coincident count rate is about  $2.6 \times 10^{4}/$s.
 Two polarizers are placed in the arms of the second entanglement pair in order to prepare the required single-photon source.
 \textbf{b}. Setups for projecting the spatial qubits onto $\ket{\alpha_{\pm}}_k = (\ket{0}_k\pm e^{i
\alpha}\ket{1}_k)/\sqrt{2}$, $H^{'}$ and $V^{'}$. \textbf{c}.
Ultra-stable Sagnac single-photon interferometer. Details are
discussed in the text.}
\end{figure}

{\em Preparation.---}The schematic setup for preparing the 4-photon
6-qubit cluster state is depicted in Fig.~\ref{Fig 2}. We use
spontaneous down conversion to produce the desired 4 photons. With
the help of polarizing beam splitters (PBSs), half-wave plates
(HWPs), and conventional photon detectors, we prepare a 4-qubit
cluster state
\begin{equation}\label{1}
\begin{gathered}
\left| {C_4 } \right\rangle = \frac{1}
{{ 2 }}[\left| + \right\rangle_1 \left| H \right\rangle_3 (\left| H \right\rangle_2 \left| + \right\rangle_4 + \left| V \right\rangle_2 \left| - \right\rangle_4 ) \hfill \\
+ \left| - \right\rangle_1 \left| V \right\rangle_3 (\left| H \right\rangle_2 \left| + \right\rangle_4 - \left| V \right\rangle_2 \left| - \right\rangle_4 )], \hfill \\
\end{gathered}
\end{equation}
where $\left|H\right\rangle$ ($\left|V\right\rangle$) represents the
state with the horizontal (vertical) polarization and $\left| \pm
\right\rangle = 1/\sqrt{2}(\left|H \right\rangle\pm\left| V
\right\rangle)$. The scheme for preparing $\left| {C_4 }
\right\rangle$ is similar to the one introduced in \cite{yama}.
After creating $\left| {C_4 } \right\rangle$, we place two PBSs in
the outputs of photons 1 and 4, as depicted in Fig.~2a. Since a PBS
transmits $H$ and reflects $V$ polarization, $H$-polarized photons
will follow one path and $V$-polarized photons will follow the
other.
In this way, the spatial qubits are added onto the polarization
qubits: $\alpha |H\rangle {}_1 + \beta |V\rangle _1  \to \alpha
|HH'\rangle {}_1 + \beta |VV'\rangle _1$, with the levels denoted as
$|H'\rangle$ for the first path and $|V'\rangle$ for the latter path
(see Fig. 2a). This process is equivalent to a controlled-phase gate
between the polarization qubit and a spatial qubit $\frac{1}{{\sqrt
2 }}(|H'\rangle  + |V'\rangle  )$ up to single-qubit unitary
transformation.

If we consider $\left| H' \right\rangle_{1,4}$ as $\left| 0
\right\rangle_{5,6}$, $\left| V' \right\rangle_{1,4}$ as $\left| 1
\right\rangle_{5,6}$ and $\left| H \right\rangle \leftrightarrow
\left| 0 \right\rangle$, $\left| V \right\rangle \leftrightarrow
\left| 1 \right\rangle$, the state will be expressed as
\begin{equation}\label{4}
\begin{gathered}
 \left| \widetilde{{LC_6 }} \right\rangle = \frac{1}{{2\sqrt 2 }}[(\left| 0 \right\rangle_5 \left| 0 \right\rangle_1 + \left| 1 \right\rangle_5 \left| 1 \right\rangle_1 )\left| 0 \right\rangle_3 \hfill \\
 (\left| \tilde{0} \right\rangle_2 \left| 0 \right\rangle_4 \left| 0 \right\rangle_6 + \left| \tilde{1} \right\rangle_2 \left| 1 \right\rangle_4 \left| 1 \right\rangle_6 ) \hfill \\
 + (\left| 0 \right\rangle_5 \left| 0 \right\rangle_1 - \left| 1 \right\rangle_5 \left| 1 \right\rangle_1 )\left| 1 \right\rangle_3 \hfill \\
 (\left| \tilde{1} \right\rangle_2 \left| 0 \right\rangle_4 \left| 0 \right\rangle_6 + \left| \tilde{0} \right\rangle_2 \left| 1 \right\rangle_4 \left| 1 \right\rangle_6 ) ], \hfill \\
\end{gathered}
\end{equation}
where
$\left|\tilde{0}\right\rangle=\frac{1}{\sqrt{2}}(\left| 0
\right\rangle+\left| 1 \right\rangle), \left| \tilde{1}
\right\rangle = \frac{1}{\sqrt{2}}(\left| 0 \right\rangle - \left| 1
\right\rangle)$. State (\ref{4}) is equivalent to a 6-qubit linear
cluster state up to two single-qubit Hadamard transformations, $H_5$
and $H_6$.




To implement the required measurements of one-way quantum
computation and estimate the fidelity of the state, we need to
project the spatial qubits onto $\ket{\alpha_{\pm}}_k = (\ket{0}_k
\pm e^{i \alpha}\ket{1}_k)/\sqrt{2}$. The required devices are shown
in Fig.~2b. When $\alpha\neq0$, the measurements are performed by
matching different spatial modes on a common BS, and the phase is
determined by the difference between the optical path length of two
input modes. Here single-photon interferometers are required in the
experiment. To achieve a high stability for the single-photon
interferometer, we have constructed an ultra-stable Sagnac setup
\cite{Nagata,Almeida} (see Fig.~2c), which can be stable for almost
10 hours \cite{Gao}. We have first designed a special crystal
combining a PBS and a beam splitter (BS). When an input photon
enters the interferometer, it is split by the PBS. The $H$ component
of the photon is transmitted and propagates counterclockwise through
the interferometer; the $V$ component is reflected and propagates
clockwise. The two spatial modes match at the BS and the
interference occurs there. After being detected by two detectors
$D_a$ and $D_b$, the output states are respectively projected onto
$\frac{1}{{\sqrt 2 }}(\left| 0 \right\rangle + e^{i\alpha } \left| 1
\right\rangle )$ and $\frac{1}{{\sqrt 2 }}(\left| 0 \right\rangle -
e^{i\alpha } \left| 1 \right\rangle )$.


{\em Fidelity.---}To characterize the quality of the generated
state, we estimate its fidelity $F = \left\langle{\widetilde{LC_6}
}\right|\rho_{exp}\left| \widetilde{LC_6} \right\rangle$. $F$ is
equal to 1 for an ideal state and $1/64$ for a completely mixed
state. We consider an observable $B$ with the property that ${\rm
tr}(B\rho_{exp}) \leq {\rm tr}(\left| \widetilde{LC_6}
\right\rangle\left\langle \widetilde{LC_6} \right|\rho_{exp}) = F$,
which implies that the lower bound of the fidelity can be obtained
by measuring observable $B$. Using the method introduced in
\cite{otfried2}, we construct the observable $B$ as
\begin{eqnarray}
&&B= \frac{1}{{32}}\{ 4(P_{51}^ - X_3 Z_2 P_{46}^ + + P_{51}^ - Y_3
Y_2
P_{46}^-)+ 2[P_{51}^- \notag\\
&&(X_3 \openone_2 X_4 X_6 - Y_3 X_2 Y_4 X_6 - X_3 \openone_2 Y_4 Y_6 - Y_3 X_2 X_4 Y_6 )]\notag\\
&&+ (X_5 Y_1 + Y_5 X_1 )[2(Y_3 Z_2 P_{46}^+ - X_3 Y_2 P_{46}^- )+ \notag \\
&&(Y_3 \openone_2 X_4 X_6 - Y_3 \openone_2 Y_4 Y_6 + X_3 X_2 X_4 Y_6 + X_3 X_2 Y_4 X_6 )]\}, \notag \\
\end{eqnarray}
where $ P_{i,j}^ \pm = \left| {00} \right\rangle
_{ij}\left\langle{00}\right| \pm \left| {11} \right\rangle
_{ij}\left\langle{11}\right|$. Experimental values of the required
measurement settings are given in Table I, from which we obtain
\begin{equation}\label{8}
F \geq {\rm tr}(B\rho_{\exp } ) = 0.61 \pm 0.01,
\end{equation}
which is clearly higher than 0.5, and thus proves the existence of
genuine 6-qubit entanglement in our state \cite{fidelity}.


\begin{table}[t]
\begin{tabular}{cccc}
\hline\hline
Observable &Value &Observable &Value\\
\hline
$X_{5}Y_{1}Y_{3}\openone_{2}X_{4}X_{6}$& $0.58\pm0.04$ & $X_{5}Y_{1}Y_{3}\openone_{2}Y_{4}Y_{6}$ & $-0.63\pm0.04$\\
$X_{5}Y_{1}X_{3}X_{2}Y_{4}X_{6}$& $0.58\pm0.04$ & $X_{5}Y_{1}X_{3}X_{2}X_{4}Y_{6}$ & $0.60\pm0.04$\\
$Y_{5}X_{1}Y_{3}\openone_{2}X_{4}X_{6}$& $0.55\pm0.04$ & $Y_{5}X_{1}Y_{3}\openone_{2}Y_{4}Y_{6}$ & $-0.56\pm0.04$\\
$Y_{5}X_{1}X_{3}X_{2}Y_{4}X_{6}$& $0.57\pm0.04$ & $Y_{5}X_{1}X_{3}X_{2}X_{4}Y_{6}$ & $ 0.60\pm0.04$\\
$P_{5,1}^ {-} X_{3}Z_{2}P_{4,6}^ {+}$& $0.64\pm0.04$& $P_{5,1}^ {-}Y_{3}Y_{2}P_{4,6}^ {-}$& $0.65\pm0.04$ \\
$P_{5,1}^ {-}X_{3}\openone_{2}X_{4}X_{6}$& $0.58\pm0.03$ &
$P_{5,1}^ {-}Y_{3}X_{2}Y_{4}X_{6}$& $ -0.66\pm0.04$ \\
$P_{5,1}^ {-}X_{3}\openone_{2}Y_{4}Y_{6}$& $-0.57\pm0.04$ &
$P_{5,1}^ {-}Y_{3}X_{2}X_{4}Y_{6}$& $-0.58\pm0.05$ \\
$X_{5}Y_{1}Y_{3}Z_{2}P_{4,6}^ {+}$& $0.57\pm0.05$ &
$X_{5}Y_{1}X_{3}Y_{2}P_{4,6}^ {-}$& $-0.65\pm0.04$ \\
$Y_{5}X_{1}Y_{3}Z_{2}P_{4,6}^ {+}$& $0.67\pm0.03$ &
$Y_{5}X_{1}X_{3}Y_{2}P_{4,6}^ {-}$& $-0.58\pm0.05$ \\
\hline\hline
\end{tabular}
\caption{Experimental values of the observables for the fidelity
estimation of $\left| {\widetilde{LC_6} } \right\rangle$. Each
experimental value is obtained by measuring in 400 seconds and
propagated Poissonian statistics of raw detection events are
considered.} \label{witnessmeasurent}
\end{table}

%



\begin{figure}[h]
 \includegraphics[width=8cm]{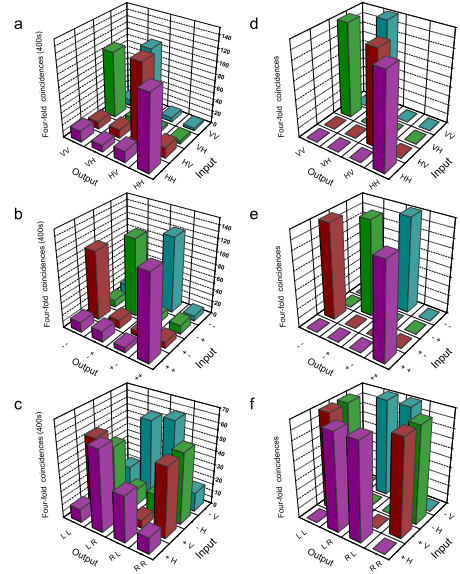}\\
 \caption{\label{Fig 3}Experimental evaluation of the process fidelity of the CNOT gate.
 In the experiment, each data is measured in 400 s.
 \textbf{a}. The experimental data of $F_{zz}$,
 defined in the basis ($\left|H \right\rangle/\left|V \right\rangle$).
\textbf{b}. The experimental values of $F_{xx}$, defined in the
basis ($\left|+ \right\rangle/\left|- \right\rangle$). \textbf{c}.
The experimental values of $F_{xz}$. The input control qubit is in
the basis ($\left|+ \right\rangle/\left|- \right\rangle$), and the
input target qubit is in the basis ($\left|H \right\rangle/\left|V
\right\rangle$), while the output qubits are measured in the basis
($\left|R \right\rangle/\left|L \right\rangle$). \textbf{d}. The
theoretical data of $F_{zz}$. \textbf{e}. The theoretical data of
$F_{xx}$. \textbf{f}. The theoretical data of $F_{xz}$. }
\end{figure}

{\em Entangling capability.---}To evaluate the performance of the
CNOT gate, we obtain the upper and lower bound of the quantum
process fidelity $F_{\rm process}$. As discussed in
\cite{Hofmann01}, $F_{\rm process}$ can be estimated as
\begin{equation}
F_{zz} + F_{xx} - 1 \le F_{\rm process} \le \min (F_{zz} ,F_{xx}),
\end{equation}
where the fidelities are defined as
\begin{eqnarray}
 F_{zz} &=& 1/4[P(HH|HH) + P(HV|HV)\notag\\
&&+ P(VV|VH) + P(VH|VV)], \notag\\
 F_{xx} &=& 1/4[P( + + | + + ) + P( - - | + - ) \notag\\
&&+ P( - + | - + ) + P( + - | - - )],
\end{eqnarray}
where each $P$ represents the probability of obtaining the
corresponding output state under the specified input state.
Experimentally, when $\{\alpha', \beta', \alpha, \beta\}$ take the
values $\{\pm\pi/2, \pm\pi/2, (0,\pi), (0,\pi)\}$, both the control
and target input qubit will lie on the basis $|H\rangle/|V\rangle$,
and when $\{\alpha', \beta',\alpha, \beta\}$ take the values
$\{(0,\pi), (0,\pi), \pm\pi/2, \pm\pi/2\}$, they will lie on the
basis $|+\rangle/|-\rangle$. The results are depicted in
Fig.~\ref{Fig 3}. $F_{zz}$ ($F_{xx}$) is $79\% \pm 2\%$ ($78\% \pm
2\%$); thus the fidelity of the gate lies between $57\% \pm 3\%$ and
$78\% \pm 2\%$.

Since the fidelity of entanglement generation is at least equal to
the process fidelity, the lower bound of the process fidelity
defines a lower bound of the entanglement capability of the gate
\cite{Hofmann01}. In terms of the concurrence $C$ which the gate can
generate from product state inputs, the minimal entanglement
capability of the gate is given by
\begin{equation}
 C \geq 2F_{\rm process}-1 \geq 2(F_{zz} +
F_{xx})-3.
\end{equation}
In our experiment, the obtained lower bound of $C$ is $0.14 \pm
0.05$, confirming the entanglement capability of our gate.


{\em Quantum parallelism.---} It was shown that a quantum CNOT gate
is capable of simultaneously performing the logical functions of
three distinct conditional local operations, each of which can be
verified by measuring a corresponding truth table of four local
inputs and four local outputs \cite{Hofmann02}. If the experimental
gate can effectively perform more than one local operation in
parallel, it is called that quantum parallelism is achieved, which
also means that the gate can't be reproduced by such local
operations and classical communication \cite{Hofmann02}. Specially,
quantum parallelism will be achieved if the average fidelity of
these three distinct conditional local operations exceeds $2/3$,
where $F_{zz}$, $F_{xx}$ are two of them, and the third one is
\begin{eqnarray}\label{444}
 F_{xz} &=& 1/4[P(RL/ + H) + P(LR/ + H) + P(RR/ + V) \notag\\
&&+ P(LL/ + V) + P(RR/ - H) + P(LL/ - H)\notag \\
&&+ P(RL/ - V) + P(LR/ - V)],
\end{eqnarray}
where $\left| R \right\rangle = 1/\sqrt{2}(\left|H
\right\rangle+i\left| V \right\rangle), \left| L \right\rangle =
1/\sqrt{2}(\left|H \right\rangle-i\left| V \right\rangle)$. $F_{xz}$
is calculated to be $80\% \pm 2\%$ (see Fig. 3c), so the average
fidelity of the three results is $79\% \pm 1\%$, obviously exceeding
the boundary 2/3 and thus proving quantum parallelism in our gate.

The imperfection of the fidelity is mainly caused by the noise in
the state generation and the imperfect interferometers. Moreover,
note that, in Hofmann's theoretical scheme of process estimation,
the input states of the tested gate should be perfect
\cite{Hofmann01, Hofmann02}, while our initial input qubits are
non-ideal due to the imperfection of the experimental cluster state,
which will  affect the accuracy of the process estimation to some
extent.




{\em Conclusion and discussion.---} In our experiment, we have
generated a four-photon six-qubit cluster state entangled in the
photons' polarization and spatial modes. In order to create new
types of cluster states and perform new one-way quantum
computations, our method  can be extended to more photons by
increasing the power of pump light \cite{Lu} or to more degrees of
freedom \cite{Kwiat2}. With the latter approach, the complexity of
the measurement apparatus may increase.

Based on the six-qubit state, we have given a  proof-of-principle
demonstration of one-way quantum CNOT gate with arbitrary
single-qubit inputs. Our results show that photons' polarization and
spatial degrees of freedom are both promising resources for
efficient optical quantum computation. As a general procedure for
application, we can first generate a cluster state entangled in
photons' polarization modes. Then, extra spatial qubits can be
planted onto the polarization qubits. The additional spatial qubits
can be used to perform local rotations, as shown in our experiment.
More recently, it is shown that by making use of additional degrees
of freedom, generalized quantum measurements (POVM) instead of
projective measurements can largely extend the quantum computational
power of cluster states \cite{ttt}. It remains an open question how
to most efficiently exploiting different degrees of freedom of
photons for quantum computation.

Finally, we would like to note that, we didn't use active
feed-forward operation in the present experiment, and thus for each
measurement of qubits, this reduced the success rate of the
computation by a factor of two compared to deterministic gate
operations. However, this suffices for a proof-of-principle
demonstration. Feed-forward operations have been developed first in
ref. \cite{Prevedel,Matini2}. By making use of delay fibers and
Pockels cells driven by the output signals of detectors, one can
efficiently perform the feed-forward operations, which can be
readily combined with our experiment in the future.


This work is supported by the NNSF of China, the CAS, the National
Fundamental Research Program (under Grant No. 2006CB921900), the FWF
(START prize), the EU (OLAQUI, QICS, SCALA), the MCI Project No.
FIS2008-05596, and the Junta de Andaluc\'{\i}a Excellence Project
No. P06-FQM-02243.


\end{document}